\documentstyle›12pt!{article}
\title{DYNAMICAL SYSTEMS AND\\ FACTORS OF FINITE AUTOMATA}
   \author{Petr K\accent23urka \vspace{8pt}\\
   Dept. of Mathematical Logic and Philosophy of Mathematics,\\
   Faculty of Mathematics and Physics, Charles University\\
Malostransk\'{e} n\'{a}m\v{e}st\'{\i} 25, 118 00 Praha 1,
Czechia\\
   {\small fax: (422) 532 742, email: kurka@cspguk11} }
\date{}

\newcommand{\mez}{\vspace{10pt}}
\newcommand{\arleft}›2! {\begin{picture}(110,10)
\put(10,0){$#2$} \put(27,3){\vector(1,0){60}} \put(50,10){$#1$}
\put(95,0){$#2$} \end{picture} }
\newcommand{\arup}›2! {\begin{picture}(110,25)
\put(3,12){$#1$}      \put(106,12){$#2$}
\put(15,2){\vector(0,1){23}} \put(100,2){\vector(0,1){23}}
\end{picture} }
\newcommand{\ardown}›2! {\begin{picture}(110,25)
\put(3,12){$#1$}      \put(106,12){$#2$}
\put(15,25){\vector(0,-1){23}} \put(100,25){\vector(0,-1){23}}
\end{picture}}

\newtheorem{dfn}{Definition}
\newtheorem{lem}{Lemma}
\newtheorem{pro}{Proposition}
\newtheorem{thm}{Theorem}
\newtheorem{cor}{Corollary}
\newtheorem{exm}{Example}

\begin{document}
\maketitle

\begin{abstract}
We conceive finite automata as dynamical systems on discontinuum
and investigate their factors. Factors of finite automata include
many well-known  simple dynamical systems, e.g. hyperbolic
systems and systems with finite attractors. In the quadratic
family on the real interval, factors of finite automata include
all systems with finite number of periodic points, as well as
systems at the band-merging bifurcations. On the other hand the
system with a non-chaotic attractor, which occurs at the limit of
the period doubling bifurcations (at the edge of chaos), is not
of this class. Next we propose as another simplicity criterion
for dynamical systems the property of having chaotic limits
(every point belongs to a set, whose $\omega$-limit is
chaotic). We show that any factor of a finite automaton has
chaotic limits but not vice versa. \mez

\noindent
key words: edge of chaos, chaotic limits, discontinuum, factors
of finite automata.
\end{abstract}

\section{Introduction}

Studies in chaotic dynamics suggest that chaotic dynamical
systems are usually rather simple, the most notable example being
the shift map. Another class of simple dynamical systems is
distinguished by the existence of periodic orbits, which attract
all points.  Systems with really complex and interesting
dynamics are found at the edge of chaos (see Langton
\cite{kn:Langton}). They are characterized by long transient
periods, strong dependence on initial conditions and the power of
universal computations. In the quadratic family on the real
interval, the systems at both period doubling and band merging
bifurcations have been shown by Crutchfield and Young
\cite{kn:Crutch} to generate simple regular languages. On the
other hand the system at their common limit generates a
non-regular language.

The feeling that chaotic means simple has been supported by a
recent result of Brooks et al. \cite{kn:Brooks}. A standard
Devaney's \cite{kn:Devaney} definition reads that a chaotic
system is one, which is topologically transitive, whose periodic
points are dense, and which is sensitive to initial conditions.
Brooks et al. \cite{kn:Brooks} prove that on infinite metric
spaces, the first two conditions imply the third. In light of
this result it is natural to relax the definition of chaotic
system to topological transitivity and density of periodic
points. According to this relaxed definition, finite dynamical
systems, which are orbits of periodic points, are chaotic too.

One might try to characterize simple systems by the condition,
that all points have chaotic $\omega$-limit sets. This would not
do, however, since even in the shift system, there are many
complex subshifts like Morse sequences. We define therefore a
system with chaotic limits as one, whose each point is included
in a set, whose $\omega$-limit set is chaotic. This works well in
the quadratic family: all systems at the period doubling and band
merging bifurcations have chaotic limits, while the system at
their common limit has not.

The simplest systems with chaotic limits are the shift map
$\sigma$ on $2^{N}$ defined by $\sigma(u)_{i}=u_{i+1}$, which
itself is chaotic, and its inverse $\sigma^{-1}_{0}$ defined by
$\sigma^{-1}_{0}(u)_{i+1} = u_{i}$,\ $\sigma^{-1}_{0}(u)_{0} =
0$, whose limit is the fixed point 0. Both these systems are
instances of finite automata. We conceive a finite automaton
as a dynamical system:
a finite control connected to infinite input and output tapes.
At each time step the automaton reads a symbol from the input
tape and writes a symbol to the output tape. The state of the
whole system consists of the inner state of the control and the
content of both tapes. The shift map is an automaton which works only
with the input tape, while its inverse works only with the output
tape. The state space of a finite automaton is a discontinuum, i.e.
a countable product of finite spaces.

Another simplicity criterion for dynamical systems is
based on the fact, that every dynamical system on a
compact metric space is a factor of a dynamical system on
discontinuum (see Balcar and Simon (\cite{kn:Balcar},
Proposition 5.12). In computer science there is a hierarchy ranging
from  finite automata, stack automata, Turing automata, and
cellular  automata to neural networks. All these automata can be
regarded as dynamical systems on discontinuum. Considering their
factors, we obtain a hierarchy of dynamical systems on compact
metric spaces. Factors of finite automata are at the bottom of
this hierarchy. This works again in the quadratic family: all
systems at the period doubling and band merging bifurcations are
factors of finite automata, while the system at their common
limit is not. We show that every factor of a finite automaton has
chaotic limits, while there are systems with chaotic limits,
which are not factors of finite automata.

In Crutchfield and Young \cite{kn:Crutch} and Troll
\cite{kn:Troll}, the complexity of dynamical systems has been
studied via complexity of languages generated by finite partitions.
Our concepts of systems with chaotic limits, and factors of finite
automata offer an alternative and complementary approach.

\section{Systems with chaotic limits}

\begin{dfn} A dynamical system $(X,f)$ is a continuous map
$f: X \rightarrow X$ of a compact metric space $X$ to itself. A
homomorphism $\varphi: (X,f) \rightarrow (Y,g)$ of dynamical
systems is a continuous mapping $\varphi:X \rightarrow Y$ such
that $g \varphi = \varphi f$. We say that $(Y,g)$ is a factor of
$(X,f)$, if $\varphi$ is a factorization (i.e. a surjective map).

\begin{picture}(200,55)
\put(110,45){\arleft{f}{X}}
\put(110,15){\ardown{\varphi}{\varphi}}
\put(110,5) {\arleft{g}{Y}}
\end{picture}
\end{dfn}

\noindent
The $n$-th iteration $f^{n}:X \rightarrow X$ of $f$ is defined by
$f^{0}(x)=x$, $f^{n+1}(x)=f(f^{n}(x))$. The orbit ${\cal O}(x) =
\{f^{i}(x)º i \geq 0\}$ of $x$ is the set of its iterates. A
point $x \in X$ is periodic with period $n>0$ if $f^{n}(x)=x$ and
$f^{i}(x) \neq x$ for $0<i<n$. A fixed point is a periodic point
with period 1. A subset $A \subseteq X$ is invariant, if $f(A)
\subseteq  A$. It is strongly invariant, if $f(A)=A$. The
$\omega$-limit set of a set $A \subseteq  X$ is defined by
$\omega(A)= \bigcap_{n} \overline{ \bigcup_{m>n} f^{m}(A)}$.  It
is easy to see that if $A$ is nonempty, then $\omega(A)$ is a
nonempty, closed, strongly invariant set, so $f$ restricted to
$\omega(A)$ is a dynamical system.

\begin{lem}
Let $\varphi: (X,f) \rightarrow (Y,g)$ be a homomorphism, and $A
\subseteq X$. Then $\varphi(\omega_{f}(A)) =
\omega_{g}(\varphi(A))$.
\end{lem} \label{omega}
Proof: If $x \in \omega_{f}(A)$, then there exist $x_{i} \in A$
and  $n_{i}$ with $f^{n_{i}}(x_{i}) \rightarrow x$, so
$\varphi(x_{i}) \in \varphi(A), g^{n_{i}}(\varphi(x_{i}))
\rightarrow \varphi(x)$, and therefore $\varphi(x) \in
\omega_{g}(\varphi(A))$. If $y \in \omega_{g}(\varphi(A))$, then
there exist $x_{k} \in A$ and $n_{k}$  with $g^{n_{k}}
\varphi(x_{k}) \rightarrow y$. There exists a convergent
subsequence $f^{n_{k_{i}}}(x_{k_{i}}) \rightarrow x \in
\omega_{f}(A)$, so $\varphi f^{n_{k_{i}}}(x_{k_{i}})
\rightarrow \varphi(x)$. It follows $y = \varphi(x)$, and
therefore $y \in \varphi(\omega_{f}(A))$. $\Box$

\begin{dfn}
A dynamical system $(X,f)$ is chaotic, if the set of its periodic
points is dense in $X$ and if for any open nonempty sets $U,V
\subseteq  X$ there exists $k>0$ such that $f^{k}(U) \cap V \neq
0$ (topological transitivity). The system $(X,f)$ has chaotic
limits, if for each $x \in X$ there exists $A \subseteq X$ such
that $x \in A$ and $\omega(A)$ is chaotic.
\end{dfn}

\noindent
In particular a finite system is chaotic iff it is an orbit of a
periodic point, and any finite system has chaotic limits. If
$\omega(x)$ is a periodic orbit for each $x \in X$, then $(X,f)$
is a system with chaotic limits. Moreover if $X$ or $\omega(X)$
is chaotic, then $(X,f)$ has again chaotic limits.

\begin{pro}
A factor of a chaotic dynamical system is chaotic. A factor of a
system with chaotic limits has chaotic limits.
\end{pro}
Proof: Suppose that  $\varphi: (X,f) \rightarrow (Y,g)$ is a
factorization and  $(X,f)$ is chaotic. Let $U \subseteq Y$ be a
nonempty open set. Then $\varphi^{-1}(U)$ is a nonempty open set
and there is a periodic point $x$ in it. It follows that
$\varphi(x)$ is a periodic point in $U$. If $U,V \subseteq  Y$
are nonempty open sets, then so are $\varphi^{-1}(U)$ and
$\varphi^{-1}(V)$, so there exists $x \in f^{k} \varphi^{-1}(U)
\cap \varphi^{-1}(V)$. It follows $\varphi(x) \in g^{k}(U) \cap
V$, so $(Y,g)$ is chaotic. Suppose now that $(X,f)$ has chaotic
limits, and let $y \in Y$. There exists $x \in X$ with
$\varphi(x)=y$ and $x \in A \subseteq X$ such that $\omega(A)$ is
chaotic. It follows that $y \in \varphi(A)$ and by Lemma \ref{omega},
$\omega_{g}(\varphi(A)) = \varphi(\omega_{f}(A))$ is chaotic. $\Box$

\section{Finite automata}

\begin{dfn}
Let $A,Q,B$ be finite sets, and $Z=\{...-1,0,1,...\}$ the set of
integers. The discontinuum (over $A,Q,B$) is the space
\› X = \{u \in (A \cup Q \cup B)^{Z} º u_{i} \in A  \mbox{ for }
i>0,\; u_{0} \in Q,\; u_{i} \in B \mbox{ for } i<0 \} \!
with metric
\› \varrho(u,v) = 2^{-n} \; \mbox{ where } \;
   n = \inf\{i \geq 0º u_{i} \neq v_{i} \; \mbox{ or } \;
   u_{-i} \neq v_{-i} \} \!
\end{dfn}

\begin{pro}
Let $X$ be a discontinuum. Then $X$ is a compact metric space.
For each compact metric space $Y$ there is a factorization
$\varphi : X \rightarrow Y$. For every dynamical system $(Y,g)$
there exists a dynamical system $f$ on $X$ and a factorization
$\varphi : (X,f) \rightarrow (Y,g)$.
\end{pro}
See Balcar and Simon \cite{kn:Balcar} for a proof.

\begin{dfn}
Let $A,Q,B$ be finite sets (input alphabet, set of inner states,
and output alphabet), and $f_{A}:Q \times A \rightarrow Q$,
$f_{B}:Q \rightarrow B$ input and output functions. A finite
automaton over $A,Q,B$ is a dynamical  system $f$ on the
discontinuum $X$ (over $A,Q,B$) defined by
\› f(u)_{-1} = f_{B}(u_{0}),\; f(u)_{0} = f_{A}(u_{0},u_{1}),\;
  f(u)_{i} = u_{i+1} \; \mbox{ for }\; -1 \neq i \neq 0 \!
\end{dfn}

\noindent
Thus $\{u_{i}ºi>0\}$ is the input tape, $\{u_{i}ºi<0\}$ is the
output tape, and\\
$f(...u_{-2},u_{-1},u_{0},u_{1},u_{2}...) =
(...u_{-1},f_{B}(u_{0}),f_{A}(u_{0},u_{1}),u_{2},u_{3}...)$.

\begin{thm}
Every finite automaton has chaotic limits.
\end{thm}
Proof: Let $w \in X$ and define $Q_{0}= \{q \in Qº (\forall
m)(\exists n >m)(f^{n}(w)_{0} = q) \}$. Let $k$ be the first
integer such that for all $n \geq k$ there is $f^{n}(w)_{0} \in
Q_{0}$. Define $W=\{v \in Xº (\forall i \leq k)(v_{i}=w_{i})\;\;
\& \;\;(\forall i \geq k)(f^{i}(v)_{0} \in Q_{0}) \}$. We shall
prove that $\omega(W)$ is chaotic. Denote $Y = \{v \in X º
(\forall n \geq 0)(f^{n}(v)_{0} \in Q_{0})\}$. Then $Y$ is an
invariant set and $\omega(W) \subseteq \bigcap \{f^{n}(Y)º n \geq
0\}$ since $f^{k}(W) \subseteq  Y$. Let $u \in \bigcap
\{f^{n}(Y)º n \geq 0\}$, and pick some $n>0$. There is $v \in Y$
with $f^{n}(v)=u$, and there exists $m \geq k$ with $f^{m}(w)_{0}
=v_{0}$. Define $s \in W$ by $s_{i}=w_{i}$ for $i \leq m$, $s_{i}
= v_{m+i}$ for $i > m$. Then $\varrho(f^{m+n}(s),u) <  2^{-n}$,
so $u \in \omega(W)$. Thus $\omega(W) = \bigcap \{f^{n}(Y)  ºn
\geq 0\}$. Let $U \subseteq  X$ be an open set and $u \in U \cap
\omega(W)$. We shall find a periodic point in $U \cap \omega(W)$.
There exists $n$ such that the $n$-th cylinder around $u$ is
contained in $U$:  $u \in \{v \in \omega(W)º (\forall ºiº \leq
n)(v_{i}=u_{i}) \} \subseteq  U \cap \omega(W)$. There exists $u'
\in Y$ such that $u=f^{n}(u')$. Then there exists $u'' \in Y$
such that $u'_{i}=u''_{i}$ for $0 \leq i \leq 2n$ and
$f^{m}(u'')_{0} = u''_{0}$ for some $m>2n$. Define $s \in X$ by
$s_{0}=u''_{0}$, $s_{km+i}=u''_{i}$, $s_{-km-i}=f^{m}(u'')_{-i}$
for $1 \leq i \leq m$, and $k \geq 0$. Then $s \in \omega(W)$ is
a periodic point, and $f^{n}(s)$ is a periodic point in $U \cap
\omega(W)$. Thus periodic points are dense in $\omega(W)$. To
prove the topological transitivity, let $U,V$ be open sets,  $u
\in U \cap \omega(W)$, $v \in V \cap \omega(W)$. There is again
$n$ such that the $n$-th cylinders around $u$ and $v$ are
contained in $U$ and $V$ respectively. Then there are $u',v' \in
Y$ such that $u=f^{n}(u')$, $v=f^{n}(v')$. There exists $u'' \in
Y$ such that $u'_{i} = u''_{i}$ for $0 \leq i \leq 2n$, and
$f^{m}(u'')_{0} =v'_{0}$ for some $m >2n$. Define $s \in X$ by
$s_{i}=u''_{i}$ for $i \leq m$ and $s_{i}=v'_{i-m}$  for $i > m$.
Then $f^{m+n}(s) \in f^{m}(U) \cap V \cap \omega(W)$. $\Box$

\begin{cor}
A factor of a finite automaton has chaotic limits.
\end{cor}

\begin{pro}
A subshift of finite type is a factor of a finite automaton.
\end{pro}
Proof: Let $A$ be a finite alphabet and $M$ a square $A \times A$
matrix with entries from $2=\{0,1\}$. Let $Q=A$, $B=\{0\}$, and
define $f_{A}$ so that for each $a,b \in A$ $(\exists c \in
A)(f_{A}(a,c)=b) \Leftrightarrow M(a,b)=1$. Define $\Sigma_{M} =
\{u \in A^{N}º (\forall i \geq 0)(M(u_{i},u_{i+1})=1) \}$,
$\sigma: \Sigma_{M} \rightarrow \Sigma_{M}$ by $\sigma(u)_{i} =
u_{i+1}$, and
$\varphi:X \rightarrow \Sigma_{M}$ by $\varphi(u)_{i} =
(f^{i}(u))_{0}$. Then $\varphi : (X,f) \rightarrow
(\Sigma_{M},\sigma)$ is a factorization. $\Box$ \mez

\noindent
Since a closed hyperbolic set with a local product structure (see
Shub \cite{kn:Shub}) is a factor of a subshift of finite type, it
is also a factor of a finite automaton.

\begin{thm} \label{attractor}
Let $(Y,g)$ be a dynamical system with finite $\omega(Y)$. Then
$(Y,g)$ is a factor of a finite automaton.
\end{thm}
Proof: Since $\omega(Y)$ is strongly invariant, it consists just
of periodic points of $(Y,g)$. For $p \in \omega(Y)$ denote
$t_{p}>0$ its period. We shall prove that for every $p \in
\omega(Y)$ there exists its closed, $g^{t_{p}}$-invariant
neighbourhood, which is disjoint with $\omega(Y)-\{p\}$. Suppose
by contradiction, that it is not the case. For $A \subseteq X$,
$\varepsilon > 0$ denote $B_{\varepsilon}(A) = \{y \in Yº
(\exists z \in A)(\varrho(y,z)<\varepsilon) \}$. For each $k>0$
the closure of $\bigcup_{n \geq 0} g^{nt_{p}}(B_{1/k}(p))$ is a
closed $g^{t_{p}}$-invariant neighbourhood of $p$, and therefore
meets $\omega(Y)-\{p\}$. It follows that there exists a sequence
$y_{k} \in B_{1/k}(p)$, such that $g^{n_{k}t_{p}}(y_{k}) \in
B_{1/k}(\omega(Y)-\{p\})$ for some $n_{k}>0$. Let $W$ be a
neighbourhood of $p$ disjoint with $\omega(Y)-\{p\}$. There
exists an open neighbourhood $U$ of $p$ with $g^{t_{p}}(U)
\subseteq W$, and also an open neighbourhood $V$ of $\omega(Y)-
\{p\}$ disjoint with $W$. For each $k$ let $m_{k}<n_{k}$ be the
first integer, for which $g^{m_{k}t_{p}}(y_{k}) \not\in U$, and
therefore $g^{m_{k}t_{p}}(y_{k}) \in Y-U-V$. Since $Y-U-V$ is
compact, there is a convergent subsequence
$g^{m_{k_{i}}t_{p}}(y_{k_{i}}) \rightarrow y \in Y-U-V$, and it
follows $y \in \omega(Y)$, which is a contradiction. So we have a
closed neighbourhood $V_{p}$ of $p$ disjoint with $\omega(Y)-
\{p\}$, for which $g^{t_{p}}(V_{p}) \subseteq  V_{p}$. Let us
prove, that $W_{p} = \{y \in Yº \lim_{k \rightarrow \infty}
g^{kt_{p}}(y) = p\}$ is an open set. If $y \in W_{p}$, there is
$k>0$ for which $g^{kt_{p}}(y) \in int(V_{p})$ and there is a
neighbourhood $U$ of $y$ with $g^{kt_{p}}(U) \subseteq  V_{p}$,
so $U \subseteq W_{p}$. Since $W_{p} \cap W_{q} = 0$ for
different $p,q \in \omega(Y)$, and $\bigcup \{ W_{p}º p \in
\omega(Y)\} = Y$, $W_{p}$ are also closed. Denote $W_{p0} = W_{p}
\cap \overline{Y-g(Y)}$, $W_{pk} = g^{k_{t_{p}}}(W_{p0})$,
$W_{p\infty} = \{p\}$. Let us show that $W_{p} = \bigcup
\{W_{pk}º 0 \leq k \leq \infty \}$. Suppose by contradiction that
$y \in W_{p} - \bigcup \{W_{pk}º 0 \leq k \leq \infty \}$. Then
there exists $y_{i}$ with $g^{it_{p}}(y_{i})=y$, and $y \in
\omega(Y)$, which is a contradiction. Construct a finite
automaton $(X,f)$ as follows: $A=\{0\}$, $Q=B=\{(p,i)º p \in
\omega(Y), 0 \leq i <t_{p}\}$, $f_{A}((p,i),0)=(p,i+1 \mbox{ mod
} t_{p})$, $f_{B}(p,i)=(p,i)$. For $p \in \omega(Y)$, $k \geq 0$
define $X_{pk} = \{u \in Xº \max\{j \geq 0º u_{-j} \neq (p,i+j
\mbox{ mod } t_{p}) \} = k+1 \}$, $X_{p\infty} = \{u \in Xº
(\forall j \geq 0)(u_{-j}=(p,i+j \mbox{ mod } t_{p}) \}$. Then
$f^{k}: X_{p0} \rightarrow X_{pk}$ is a homeomorphism. For each
$p \in \omega(Y)$ $X_{p0}$ is a discontinuum, and $W_{p0}$ is a
compact space. We define $\varphi$ on $X_{p0}$ so that
$\varphi(X_{p0})=W_{p0}$. For $k>0$ define $\varphi(u) = g^{k}
\varphi f^{-k}(u)$, and for $u \in W_{p\infty}$  $\varphi(u)=p$.
Then $\varphi: (X,f) \rightarrow (Y,g)$ is a factorization.
$\Box$ \mez

\noindent
A set $A \subseteq  X$ is an attractor (see Devaney
\cite{kn:Devaney}),  if it has a neighbourhood $U$ such that
$f(U) \subseteq  int(U)$,  and $A=\omega(U)$. Thus Theorem
\ref{attractor} is concerned with the systems whose only
attractors are finite.

\begin{dfn}
A sum of dynamical systems $(X_{0},f_{0})$ and $(X_{1},f_{1})$ is
a dynamical system $(X,f)$, where $X = X_{0} \cup X_{1}$ is the
disjoint union, and  $f(x) = f_{i}(x)$ if $x \in X_{i}$.
\end{dfn}

\begin{pro} \label{sum}
The sum of dynamical systems, which are factors of finite automata,
is itself a factor of a finite automaton.
\end{pro}
Proof: Let $\varphi_{0}: (X_{0},f_{0}) \rightarrow (Y_{0},g_{0})$,
$\varphi_{1}: (X_{1},f_{1}) \rightarrow (Y_{1},g_{1})$,
be factorizations of finite automata. Define a finite automaton
$(X,f)$ by $A= A_{0} \times A_{1}$, $Q = Q_{0} \cup Q_{1}$,
(disjoint union) $B = B_{0} \times B_{1}$. Pick some $b_{0}
\in B_{0}$, $b_{1} \in B_{1}$ and define
\› \begin{array}{lll}
   f_{A}(q,(a_{0},a_{1})) = f_{A_{0}}(q,a_{0}), &\;
   f_{B}(q) = (f_{B_{0}}(q),b_{1}) & \mbox{ if }\; q \in Q_{0}\\
   f_{A}(q,(a_{0},a_{1})) = f_{A_{1}}(q,a_{1}), &\;
   f_{B}(q) = (b_{0},f_{B_{1}}(q)) & \mbox{ if }\;  q \in Q_{1}
    \end{array} \!
Let $(Y,g)$ be the sum of $(Y_{0},g_{0})$ and $(Y_{1},g_{1})$.
Define $\varphi : (X,f) \rightarrow  (Y,g)$ by
$\varphi(u) = \varphi_{0} \pi_{0}(u)$ if $u_{0} \in Q_{0}$,
$\varphi(u) = \varphi_{1} \pi_{1}(u)$ if $u_{0} \in Q_{1}$.
Here $\pi_{i}:X \rightarrow X_{i}$ are the projections.

\section{Clopen-regular systems}

If $\alpha$ is a finite set, denote $\alpha^{*}$ the set of words
over $\alpha$. For a word $u \in \alpha^{*}$ denote $º\alphaº
\geq 0$ its length and $u_{i}$ its $i+1$st letter, so $u =
u_{0}...u_{ºuº-1}$. Subsets of $\alpha^{*}$ are called languages.
See Hopcroft and Ullmann \cite{kn:Hop} for the definition of a
regular language.

\begin{dfn}
Let $(X,f)$ be a dynamical system and $\alpha$ a finite set of
subsets of $X$. Define the language generated by $\alpha$ as \›
L(f,\alpha) = \{ u \in \alpha^{*}º (\exists x \in X)(\forall i <
ºuº)(f^{i}(x) \in u_{i}) \} \!
The system $(X,f)$ is clopen-regular if $L(f,\alpha)$ is regular
for every finite clopen cover $\alpha$ of $X$.
\end{dfn}

\noindent
A subset $A$ of a metric space $X$ is clopen, if it is both
closed and open. A set $\alpha$ is a cover of $X$ if its union is
$X$. It is a partition, if it is a cover, whose different sets
are disjoint. Let $X$ be discontinuum (over $A,Q,B$), $w \in X$
and $n>0$. Then the $n$-cylinder $›w!_{n} = \{v \in X º (\forall
i \in Z)(ºiº \leq n \Rightarrow  v_{i}=w_{i}) \}$ of $w$ is a
clopen set (it is a ball with centre $w$). The set $\alpha_{n} =
\{›w!_{n}º w \in X\}$ of all $n$-cylinders is a finite clopen
partition.

\begin{pro}
Let $(X,f)$ be a dynamical system on discontinuum such that
$L(f,\alpha_{n})$ is regular for each $n>0$ ($\alpha_{n}$ is the
partition of $n$-cylinders). Then $(X,f)$ is clopen-regular.
\end{pro}
Proof: Let $\beta$ be a finite clopen cover. For each $u \in X$
there exists $n_{u}>0$ such that if $u \in B \in \beta$, then
$›u!_{n_{u}} \subseteq B$, and if $u \not \in B \in \beta$ then
$›u!_{n_{u}} \cap B = 0$. The set $\{ ›u!_{n_{u}} º u \in X \}$
is an open cover of $X$. Since $X$ is compact, there is a finite
subcover $\{›u!_{n_{u}}º u \in I\}$. Let $n= \max\{n_{u}º u \in
I\}$. Then for each $B \in \beta$ there is a finite set $J_{B}
\subseteq \alpha_{n}$ such that if $A \in J_{B}$ then $A
\subseteq B$, and if $A \not\in J_{B}$, then $A \cap B = 0$.
Since $L(f,\alpha_{n})$ is regular, $L(f,\beta)$ is regular too.
$\Box$

\begin{pro}
A factor of a clopen-regular system is clopen-regular.
\end{pro}
Proof: Let $\varphi : (X,f) \rightarrow (Y,g)$ be a factorization,
and $\alpha$ a clopen cover of $Y$. Then $\varphi^{-1}\alpha =
\{\varphi^{-1}(A)º A \in \alpha \}$ is a clopen cover of $X$
and $L(f,\varphi^{-1}\alpha) = L(g,\alpha)$. $\Box$

\begin{pro}
Any finite automaton is clopen-regular.
\end{pro}
Proof: Let $\alpha_{n}$ be the clopen partition of $n$-cylinders.
For $A,B \in \alpha_{n}$ define $M(A,B)=1$ if $f(A) \cap B \neq
0$ and $M(A,B)=0$ otherwise. Then $L(f,\alpha_{n}) = \{u \in
\alpha^{*}_{n}º (\forall i<ºuº-1) (M(u_{i},u_{i+1})=1) \}$. This
is a regular language. $\Box$

\begin{cor}
Any factor of a finite automaton is clopen-regular.
\end{cor}

\begin{exm}
There exists a chaotic dynamical system which is not a factor of
any finite automaton.
\end{exm}
Proof: Let $X=\{u \in 2^{N}º 0^{n}1^{m}0 \prec u \Rightarrow n
\leq m \}$. Here $2=\{0,1\}$ and $\prec$ means substring. Thus $u
\in X$ if every string of $n$ zeros in $u$ is immediately
followed by a string of at least $n$ ones. Then $X$ is a closed,
shift-invariant subset of $2^{N}$, and it is easy to see that
$(X,\sigma)$ is chaotic. However, it is not clopen-regular.
Consider the partition
$\alpha_{0} = \{ \{uºu_{0}=0\}, \{uºu_{0}=1\} \}$. Then
$L(\sigma,\alpha_{0}) = \{u \in 2^{*}º 0^{n}1^{m}0 \prec u
\Rightarrow n \leq m \}$ consists just of all substrings of
elements of $X$ and is not regular. $\Box$

\begin{exm}
There exists a dynamical system $(X,f)$, such that $\omega(x)$ is
a fixed point for all $x \in X$, but $(X,f)$ is not a factor of a
finite automaton.
\end{exm}
Proof: Let $M \subseteq  N = \{0\}^{*}$ be a nonregular set of
positive integers. Define $X = \{u \in 3^{Z}º i<j \Rightarrow
u_{i} \leq u_{j}\;\; \&\;\; 01^{m}2 \prec u \Rightarrow m \in M
\}$, where $3=\{0,1,2\}$.  Thus $u \in X$ if it is nondecreasing,
and when it contains a finite number of ones, then their number
belongs to $M$. Again, $X$ is a closed shift-invariant set. Its
only fixed points are constant sequences $\overline{0},
\overline{1}, \overline{2}$, and for each $x \in X$, $\omega(x)$
is one of them. However, $(X,\sigma)$ is not clopen-regular. If
$\alpha_{0}$ is the partition of the $0$-cylinders, then
$L(\sigma,\alpha_{0}) = \{u \in 3^{*}º i<j
\Rightarrow u_{i} \leq u_{j}\;\; \& \;\; 01^{m}2 \prec u
\Rightarrow m \in M \}$ is again the language of finite
substrings of elements of $X$, which is not regular. $\Box$ \mez

\noindent
This example does not contradict Theorem \ref{attractor}, since
$\sigma(X)=X$ and therefore $\omega(X)=X$ is infinite. Consider
$X_{1}=\{u \in Xº u_{1} \geq 1 \}$ Then $X_{1}$ is a
closed shift-invariant subset of $X$ with only two fixed points.
When we factorize them, we get a system with a single fixed
point, which is $\omega$-limit of all other points, and still the
system is not a factor of a finite automaton. This shows that the
Theorem \ref{attractor} cannot be substantially strengthened.

\section{Unimodal systems}

\begin{dfn} A symmetric real function $g$ defined on  $I=›-1,1!$
is unimodal, if it is continuously differentiable,
$g(-1)=g(1)=-1$, $g(0)>-1$ and $g'(x)=0$ iff $x=0$.
(We speak about unimodal system, if $g(I) \subseteq I$, i.e. if
$g(0) \leq 1$.) \end{dfn}

\begin{pro} Let $g:I \rightarrow R$ be a symmetric unimodal
function with $g(0)>0$. Then there is a unique $0<p<1$ with
$g(-p)=g(p)=p$. Define the renormalization of $g$ as ${\cal
R}g(x) = - g^{2}(px)/p$. Then ${\cal R}g: I \rightarrow  R$ is a
symmetric unimodal function.
\end{pro}
Proof: Clearly $g'(x)>0$ for $x<0$, and $g'(x)<0$ for $x>0$.
Since $g(x)-x$ is positive for $x=0$ and negative for $x=1$,
there is a unique fixed point $p>0$, and $g(-p)=p$. Clearly
${\cal R}g(\pm 1) = - g^{2}(\pm p)/p=-1$. For $x \in (-p,p)$
there is $g(x)>p$, so $g'(g(x))<0$. The equation $(g^{2})'(x)=
g'(g(x)) g'(x)=0$ has in $›-p,p!$ unique solution $0$, so the
equation $({\cal R}g)'(x)=0$ has in $›-1,1!$ the unique solution
$0$. Since $g$ is decreasing for $x>0$, and $0<p$, we have $g(0)
>p$, and $g^{2}(0)<p$. It follows ${\cal R}g(0) = -g^{2}(0)/p >
-1$. $\Box$

\begin{thm} \label{renorm}
Let $g$ be a symmetric unimodal system with $g(0)>0$. If ${\cal
R}g(0)\leq 1$, and if ${\cal R}g$ is a factor of a finite
automaton, then $g$ is also a factor of a finite automaton.
\end{thm}

\begin{picture}(200,60)
\put(65,45){\arleft{f}{X}}
\put(65,15){\ardown{\varphi}{\varphi}}
\put(65,5) {\arleft{g^{2}}{I_{0}}}
\put(205,45){\arleft{h}{X''}}
\put(205,15){\ardown{\psi}{\psi}}
\put(205,5) {\arleft{g}{I}}
\end{picture}

\noindent
Proof: Let $p_{0}$ be the positive fixed point of $g$. There is
an increasing sequence $p_{k} \in (0,1)$ such that $g(-p_{k})$ =
$g(p_{k})$ = $-p_{k-1}$ for $k>0$. Denote $I_{-\infty}=\{-1\}$,
$I_{k}=›-p_{-k},-p_{-k+1}!$ for $k<0$, $I_{0}=›-p_{0},p_{0}!$,
$I_{k}=›p_{k-1},p_{k}!$ for  $k>0$, and $I_{\infty}=\{1\}$. Then
$g(I_{-k}) = g(I_{k}) = I_{-k+1}$ for $k>0$. Since ${\cal R}g(I)
\subseteq I$, we have $g^{2}(I_{0}) \subseteq I_{0}$, and
$g(I_{0}) \subseteq I_{1}$. By the assumption there exists
a finite automaton $(X,f)$ and a factorization $\varphi:(X,f)
\rightarrow (I_{0},g^{2})$. We construct a finite automaton
$(X',h)$. Suppose $A=\{0,...,m-1\}$ and define $A'=\{0,...,m\}$,
$Q'=\{(q,i)º q \in Q, i\in \{-1,0,1,\sharp\}\}$, $B'=B$. For $x
\in Q'$ denote $\lfloor x \rfloor$ its first component, and
$\lceil x \rceil$ the second, so that $x=(\lfloor x
\rfloor,\lceil x \rceil)$. Define
\› \begin{array}{ll}
 X_{-\infty} & = \{u \in X'º\; \lceil u_{0} \rceil=-1,
       (\forall j >0)(u_{j}=m) \}, \\
 X_{k} & = \{u \in X'º\; \lceil u_{0} \rceil = \mbox{sgn}(k),
   \min\{j > 0ºu_{j} \neq m\} = ºkº \},\;\; ºkº > 0 \\
X_{\infty} & = \{u \in X'º\; \lceil u_{0} \rceil=1,  (\forall j
>0)(u_{j}=m) \}  \\
  X_{i} & = \{u \in X'º\; \lceil u_{0} \rceil = i \} \;\;
     \mbox{ for }\;\; i \in \{0,\sharp\} \end{array} \!
Define $h_{B} : Q' \rightarrow  B'$, $h_{A}: Q' \times A'
\rightarrow Q'$  by $h_{B}(x) = f_{B}(\lfloor x \rfloor)$,
\› \begin{array}{lll}
  h_{A}(x,m) & = (\lfloor x \rfloor, -1) & \mbox{ if } \;
               \lceil x \rceil \in \{-1,1\} \\
  h_{A}(x,j) & = (\lfloor x \rfloor, 0) & \mbox{ if } \;
               \lceil x \rceil \in \{-1,1,\sharp\},\; j<m \\
h_{A}(x,j) & = (f_{A}(\lfloor x \rfloor,j), \sharp) & \mbox{ if }
             \; \lceil x \rceil = 0,\; j<m \\
  h_{A}(x,m) & = h_{A}(x, m-1) & \mbox{ if }
             \; \lceil x \rceil \in \{0,\sharp\} \end{array} \!
Then $h(X_{-k}) = h(X_{k}) = X_{-k+1}$, $h(X_{0}) \subseteq
X_{\sharp}$,  $h(X_{\sharp}) = X_{0}$.  Denote
$X''=\{u \in X'º (\forall k>0)(u_{k}=m \Rightarrow \lceil u_{0}
\rceil \in \{0,\sharp\} \;\;   \&\;\; (\forall 0<j<k)(u_{j}=m ))
\}$. Thus $m$'s are allowed only at the beginning of the input tape
and only if $\lceil u_{0} \rceil \in \{-1,1\}$. $X''$ is a closed
and invariant set, and there is a factorization $\chi:(X',h)
\rightarrow (X'',h)$ defined by \\
$\chi(u)_{k} = m-1 \mbox{ if } k>0, \; u_{k}=m, \mbox{ and }
\lceil u_{0} \rceil \in \{0,\sharp\} \mbox{ or } (\exists
0<j<k)(u_{j}<m)$ \\
$\chi(u)_{k} = u_{k} $ otherwise.
We construct a projection $\psi: (X'',h) \rightarrow (I,g)$. Let
$g_{0}:›-1,g(0)! \rightarrow ›-1,0!$, $g_{1}:›-1,g(0)!
\rightarrow ›0,1!$ be the two inverse functions for $g$. Then
define by induction
\› \begin{array}{lll}
  \psi(u) = \varphi(...u_{-4}u_{-2}\lfloor u_{0} \rfloor
   u_{1}u_{3}...) &  \mbox{ if } & \lceil u_{0} \rceil =0 \\
\psi(u) = g_{0}^{k} \psi h^{k}(u) & \mbox{ if } & \lceil u_{0}
\rceil = -1 \mbox{ and } \lceil h^{k}(u)_{0} \rceil = 0 \\
      \psi(u) = g_{1} \psi h(u) & \mbox{ if } & \lceil u_{0}
\rceil  \in \{1,\sharp\}\\
  \psi(u) = -1 & \mbox{ if } & u \in X_{-\infty}\\
  \psi(u) =  1 & \mbox{ if } & u \in X_{\infty}  \end{array} \!
Then clearly $\psi h(u) = g \psi(u)$ whenever $\lceil u_{0}
\rceil \neq 0$. For $\lceil u_{0} \rceil = 0$ observe that both
$\psi h(u)$ and $g \psi(u)$ belong to $›p,1!$. Since $g$ is
decreasing on $›p,1!$ it suffices to show $g \psi h(u) = g^{2}
\psi(u)$. Since $\lceil h(u)_{0} \rceil \neq 0$, we have $g \psi
h(u)$ = $\psi h^{2}(u)$ =\\
$\psi h (... u_{-2},u_{-1},f_{B}(\lfloor u_{0} \rfloor),
   (f_{A}(\lfloor u_{0} \rfloor,u_{1}),\sharp),u_{2},u_{3}, ...)$ =\\
$\psi (... u_{-2},u_{-1},f_{B}(\lfloor u_{0} \rfloor),
f_{B}f_{A}(\lfloor u_{0} \rfloor,u_{1}),
(f_{A}(\lfloor u_{0} \rfloor,u_{1}),0), u_{3},u_{4} ...)$ =\\
$\varphi(... u_{-4},u_{-2},f_{B}(\lfloor u_{0} \rfloor),
f_{A}(\lfloor u_{0} \rfloor, u_{1}),u_{3},u_{5} ...)$ =\\
$\varphi f (...u_{-4}u_{-2} \lfloor u_{0} \rfloor u_{1}u_{3}...)$
= $g^{2} \varphi (...u_{-4}u_{-2} \lfloor u_{0} \rfloor
u_{1}u_{3}...)$ = $g^{2} \psi (u)$. $\Box$ \mez

\begin{lem} \label{increase}
Let $g : I \rightarrow  I$ be an increasing continuous function
such that $g(x) > x$ iff $x \in (-1,1)$. Then $(I,g)$ is a factor
of a finite automaton.
\end{lem}

\begin{picture}(300,60)
\put(65,45){ \begin{picture}(250,10) \put(-15,3){. . .}
\put(13,0){$\underline{0}01$} \put(32,3){\vector(1,0){40}}
\put(78,0){$\underline{0}1$}  \put(92,3){\vector(1,0){35}}
\put(137,0){$0\underline{1}$} \put(152,3){\vector(1,0){40}}
\put(196,0){$01\underline{1}$} \put(222,3){. . .}
\put(40,7){$f$} \put(105,7){$f$} \put(170,7){$f$} \end{picture} }
\put(65,15){  \begin{picture}(250,25)
  \put(3,12){$\varphi$}  \put(15,25){\vector(0,-1){23}}
 \put(68,12){$\varphi$}  \put(80,25){\vector(0,-1){23}}
\put(153,12){$\varphi$}  \put(145,25){\vector(0,-1){23}}
\put(218,12){$\varphi$}  \put(210,25){\vector(0,-1){23}}
\end{picture} }
\put(65,5){ \begin{picture}(250,10) \put(-15,3){. . .}
\put(10,0){$I_{-1}$} \put(40,7){$g$} \put(29,3){\vector(1,0){40}}
\put(75,0){$I_{0}$} \put(105,7){$g$} \put(92,3){\vector(1,0){40}}
\put(140,0){$I_{1}$} \put(170,7){$g$} \put(157,3){\vector(1,0){40}}
\put(205,0){$I_{2}$} \put(222,3){. . .}   \end{picture} }
  \end{picture}

\noindent
Proof: For $k \in Z$ denote $I_{k}=›g^{k}(0),g^{k+1}(0)!$ ($f$ is
a homeomorphism), $I_{-\infty} = \{-1\}$, $I_{\infty}=\{1\}$.
Define $A=Q=B=\{0,1\}$, $f_{B}(i)=i$, and $f_{A}(i,j) =
\max\{i,j\}$. We have closed sets
\› \begin{array}{lll}
   X_{-\infty} & = \{u \in Xº (\forall i \geq 0)(u_{i}=0) \}\\
   X_{k} & = \{u \in Xº \min\{i \geq 0º u_{i}=1\} = -k+1 \} , &
        k \leq 0 \\
   X_{k} & = \{u \in Xº \max\{i \leq 0º u_{i}=0\} = -k \} , &
      k > 0 \\
   X_{\infty} & = \{u \in Xº (\forall i \leq 0)(u_{i}=1) \}
\end{array} \!
For $k \in Z$ define a projection $\pi_{k}: X \rightarrow 2^{N}$
by $\pi_{k}(u)_{i} = 0$ if $0 \leq i < -k$, $\pi_{k}(u)_{i} =
u_{-k-i-1}$ if $i \geq \max\{0,-k\}$. (Thus $\pi_{-1}(u) =
(0,u_{-1},u_{-2}...)$, $\pi_{1}(u) = (u_{-2}u_{-3}...)$. It is
easy to see that $\pi_{k+1}f(u)=\pi_{k}(u)$ for $u \in X_{k}$.
Since $I_{0}$ is compact, there exists a factorization $\psi :
2^{N} \rightarrow I_{0}$. Define $\varphi(u) = g^{k} \psi
\pi_{k}(u)$ for $u \in X_{k}$, $\varphi(u)=-1$ for $u \in X_{-
\infty}$, and $\varphi(u) = 1$ for $u \in X_{\infty}$. Then
$\varphi: (X,f) \rightarrow (I,g)$ is a factorization. $\Box$

\begin{pro} \label{negative}
Let $(I,g)$ be a unimodal system with $g(0) \leq 0$. Then it is a
factor of a finite automaton.
\end{pro}
Proof: There is at most one fixed point in $(-1,0!$. If there is
none, we get the result by Theorem \ref{attractor}. Suppose there
is a fixed point $-1<p \leq 0$. Then $I_{0}=›-1,p!$ and
$I_{1}=›p,0!$ are invariant sets for $g$. $(I_{0},g)$ is a factor
of a finite automaton by Lemma \ref{increase}.  $(I_{1},g)$ is
a factor of a finite automaton by Theorem \ref{attractor}.
Their sum is a factor of a finite automaton by Proposition \ref{sum}.
But $(›-1,0!,g)$ is a factor of this sum, so
there is a factorization $\varphi : (X,f) \rightarrow
(›-1,0!,g)$. Define an automaton $(X',f')$ by $A'=A$, $B'=B$,
$Q'=Q \times 2$, $f'_{A}(q,i)=(f_{A}(q),0)$,  $f'_{B}(q,i) =
f_{B}(q)$. Define a factorization $\varphi':(X',f) \rightarrow
(I,g)$ by $\varphi'(u) = \varphi(...u_{-1},q,u_{1}...)$ if
$u_{0}=(q,0)$, $\varphi'(u) = g^{-1}_{1} \varphi f(u)$ if
$u_{0}=(q,1)$. $\Box$

\begin{thm}
Let $(I,g)$ be a symmetric unimodal system with a finite number
of periodic points. Then it is a factor of a finite automaton.
\end{thm}
Proof: If $g(0)>0$, then there exists ${\cal R}(g)$ and $I$ is
its invariant set (otherwise $g$ would have infinite number of
periodic points). If ${\cal R}g(0)>0$, we construct further
renormalizations. Since $g$ has only a finite number of periodic
points, there exists $k \geq 0$, with ${\cal R}^{k}g(0) \leq 0$.
By Proposition \ref{negative}, ${\cal R}^{k}g$ is a factor of a
finite automaton, and by Theorem \ref{renorm}, so is $g$. $\Box$

\begin{pro}
Let $(I,g)$ be a symmetric unimodal system with $g(0)=1$. Then
$g$ is a factor of a finite automaton.
\end{pro}
Proof: Define a closed cover $\alpha= \{›-1,0!,›0,1!\}$, and a
mapping $\varphi: 2^{N} \rightarrow I$  by
$\varphi(u) = x \in I$ iff $(\forall i \geq 0)(g^{i}(x) \in
\alpha_{i} \}$. It follows from the results in Collet and Eckmann
\cite{kn:Collet}, that $\varphi$ is a surjective mapping, and
$\varphi: (2^{N},\sigma) \rightarrow (I,g)$ is a factorization.
$\Box$

\begin{cor}
Let $(I,g)$ be a symmetric unimodal system at a band-merging
bifurcation, i.e. ${\cal R}^{k}g(0)=1$ for some $k$. Then it is
a factor of a finite automaton.
\end{cor}

\begin{exm}
A symmetric unimodal system at the common limit of period
doubling and band merging bifurcations has not chaotic limits.
\end{exm}
Proof: There exists an infinite closed invariant set $Y \subseteq
I$, which is topologically transitive but has no periodic point.
The set of periodic points $P \subseteq  I$ is infinite too, and
its closure is $P \cup Y$. Moreover, for any finite $A \subseteq P$
there exists $U \subseteq I$ with $Y \subseteq U$ and $\omega(U) \cap
A = 0$ (see Collet and Eckmann
\cite{kn:Collet} and Falconer \cite{kn:Fal} for more details).
Let $y \in Y$ and $y \in A \subseteq  I$. If $\omega(A) \subseteq Y$,
then it has no periodic point. If $\omega(A) \cap P \neq 0$,
then $\omega(A)$ is not topologically transitive. In either case
$\omega(A)$ is not chaotic.  $\Box$
\pagebreak

\end{document}